\documentclass[iop]{emulateapj}

\usepackage{epsfig}
\usepackage{graphicx}
\usepackage{dcolumn}
\usepackage{bm}
\usepackage[colorlinks=true,linkcolor=blue,citecolor=blue]{hyperref}
\usepackage{natbib}
\usepackage{comment}
\usepackage{xcolor}
\usepackage{enumerate}
\usepackage{bigints}
\usepackage{multirow}
\usepackage{booktabs} 
\usepackage{mathtools}
\usepackage{makecell,tabularx}
\usepackage[normalem]{ulem}
\setcellgapes{3pt}

\newcommand{\yr}{\, \text{yr}}

\newcommand{\Mpcmcube}{{{\, \rm Mpc}^{-3}}}
\newcommand{\Gpcmcube}{{{\, \rm Gpc}^{-3}}}
\newcommand{\yrmone}{{{\, \rm yr}^{-1}}}
\newcommand{\megayr}{{{\, \rm Myr}}}

\newcommand{\ud}{{\rm d}}

\newcommand{\msun}{{{\, \rm M}_\odot}}

\let\oldtextbf=\textbf
\renewcommand\textbf[1]{{\boldmath\oldtextbf{#1}}}

\graphicspath{{./}{figures/}}

\shorttitle{Stellar black hole binaries above the pair instability mass gap}
\shortauthors{Mangiagli et al.}

\begin{document}

\title{Merger rate of stellar black hole binaries above the pair-instability mass gap}

\author{Alberto Mangiagli$^{1,2}$, Matteo Bonetti$^{1,2}$, Alberto Sesana$^{1}$ \& Monica Colpi$^{1,2}$}
\email{a.mangiagli@campus.unimib.it}
\altaffiltext{1}{Department of Physics G. Occhialini, University of Milano - Bicocca, Piazza della Scienza 3, 20126 Milano, Italy}
\altaffiltext{2}{National Institute of Nuclear Physics INFN, Milano - Bicocca, Piazza della Scienza 3, 20126 Milano, Italy}

\begin{abstract}  
In current stellar evolutionary models, the occurrence of pair-instability supernovae implies the lack of stellar black holes (BHs) with masses between about $[60, \, 120] \msun$, resulting in the presence of an upper mass gap in the BH mass distribution.
In this Letter, we propose a simple approach to describe BHs beyond the pair-instability gap, by convolving the initial mass function and star formation rate with the metallicity evolution across cosmic time. Under the ansatz that the underlying physics of binary formation does not change beyond the gap, we then construct the cosmic population of merging BH binaries. 
The detection rate of BH binaries with both mass components above the gap is found to range between  $\simeq [0.4,\,7] \, \yrmone$ for LIGO/Virgo at design sensitivity and  $[10, \, 460] \, \yrmone$ for third-generation ground-based detectors, considering the most pessimistic and optimistic scenarios. 
LISA can individually detect these binaries up to thousands of years from coalescence. The number of events merging in less than four years, which enable multi-band observation in sequence, is expected to be in the range $[1, \, 20]$. While ET will detect all these events, LIGO/Virgo is expected to detect $\lesssim 50\%$ of them. 
Finally, we estimate that the gravitational-wave background from unresolved sources in the LISA band may in principle be detected with a signal-to-noise ratio between $ \simeq 2.5$ and $\simeq 80$.

\end{abstract}

\color{black}

\section{Introduction}
\label{sec:intro} 

During the first and second observing runs, the LIGO-Virgo scientific collaboration \citep{2015, Acernese_2014} detected the gravitational-wave (GW) signals from the coalescence of ten stellar black hole binaries (BHBs) with  individual masses between $7.7^{+2.2}_{-2.6}$ and $50.6^{+16.6}_{-10.2}\msun$  \citep{LIGOScientific:2018mvr}. The observed events can be reproduced by stellar population synthesis codes \citep[see e.g.][]{Podsiadlowski2003, Postnov2014, Dominik:2014yma, Spera_2015, Belczynski:2016obo, 10.1093/mnrasl/slx118, 10.1093/mnras/stz170} in which a key role is played by the metallicity evolution along the cosmic history. In fact, it is widely accepted that low metallicity stars experience negligible mass loss during their lifetimes, due to their weaker stellar wind, thus collapsing in heavy BH remnants, consistent with those discovered by LIGO-Virgo \citep{PhysRevLett.116.241102,PhysRevLett.118.221101}.

The occurrence of pulsational pair-instability supernovae (PPISNe) and pair-instability supernovae (PISNe) in massive, low metallicity stars \citep[with $\rm Z\lesssim 0.002$,][]{Heger2002,Woosley_2017,Yoshida_2016,Marchant2018} is expected to enhance the formation of BHs in the mass range $30 \lesssim M_{\rm{rem}}/\msun \lesssim 50$, leading to a pile up around $\sim 45\msun$ \citep{Stevenson_2019}. Current GW data indicate an excess of BHs in the interval $30-45\msun$  \citep{LIGOScientific:2018mvr}, that future observations can confirm or challenge \citep{Fishbach_2017,Talbot:2018cva}. Above $\sim 50\msun$, a cut-off or edge is expected in the BH mass function, as PISNe lead to the explosion of the star preventing the formation of a massive BH remnant. However, stars with zero-age main sequence (ZAMS) mass $M_{\rm{ZAMS}} \gtrsim 260 \msun$ and absolute metallicities $\lesssim 10^{-3}$ avoid disruption, as they develop massive CO cores that directly collapse into a BH of $M_{\rm rem} \gtrsim 100 \msun$ \citep{Woosley_2002,Uchida2019}. Those systems have been invoked as viable seeds of supermassive BHs in the high redshift universe \citep{Volonteri_2010,10.1093/mnras/stw225}, but have so far been ignored in population synthesis models used to interpret LIGO-Virgo detections, that customarily evolve stars only up to $100-150\msun$ \citep{Belczynski2016,Mapelli2019MNRAS, Neijssel2019}. Conversely, several alternative mechanisms to produce BHs above the PISNe cut-off have been proposed \citep{Gerosa17,Rodriguez_2018,2019arXiv190704356M, 2019arXiv190609281Y}. Moreover, \citealt{Spera2019} and \citealt{ 10.1093/mnras/stz1453}  proposed that BHs in the pair-instability gap may originate from  the direct collapse of massive stars with large envelope and small core masses, thus avoiding the pair-instability phase. 
 
If BHs above the PISNe ``upper-mass'' gap do indeed form, pair and coalesce in binaries, they could be potentially detectable with the Laser Interferometer Space Antenna (LISA, \citealt{2017arXiv170200786A}), or from third-generation ground-based detectors such as  the Einstein Telescope \citep[ET,][]{Punturo2010}. These binaries may also contribute to the stochastic GW background (GWB) between 0.5 mHz and 20 mHz, hampering observations of individual sources close to the LISA bucket \citep{Caprini2019}. 

In this Letter, we estimate the merger rate of stellar BHBs from isolated field binary evolution across the mass spectrum and beyond the upper-mass gap. Assuming the gap as a sharp cut-off at $[60,\,  120] \msun$, we distinguish three sub-populations for the binaries: the `above-gap' (`below-gap') binaries with both components above (below) the upper (lower) edge of the mass gap and the `across-gap' binaries  with one component above and one below the mass gap. The population of `below-gap' binaries is found to be consistent with previous studies \citep[e.g.,][]{Sesana2016,Gerosa:2019dbe} and is not considered here. For the `across-gap' and `above-gap' sub-populations, we report detection rates with ground- and space-based detectors and estimate their contribution to the stochastic GWB in the LISA band. 

\section{Models}
\label{sec:models}

Our approach builds on knowledge of the `below-gap' BHB population, extending it to BHB properties above the PISNe gap.

We evolve single stars using the stellar evolution code SEVN \citep[][and reference therein]{Spera_2017}. The code includes up-to-date stellar winds, SN explosion models, PISNe and PPISNe prescriptions, and provides BH remnant masses as a function of the mass of the progenitor stars and of the absolute metallicity, in the range $2 \times 10^{-4} < \rm{Z} < 2 \times 10^{-2}$ (see Fig. 2 in \citealt{Spera_2017} for the relation between initial stellar mass and remnant mass for different metallicities).

\begin{figure}
	\includegraphics[width=\columnwidth]{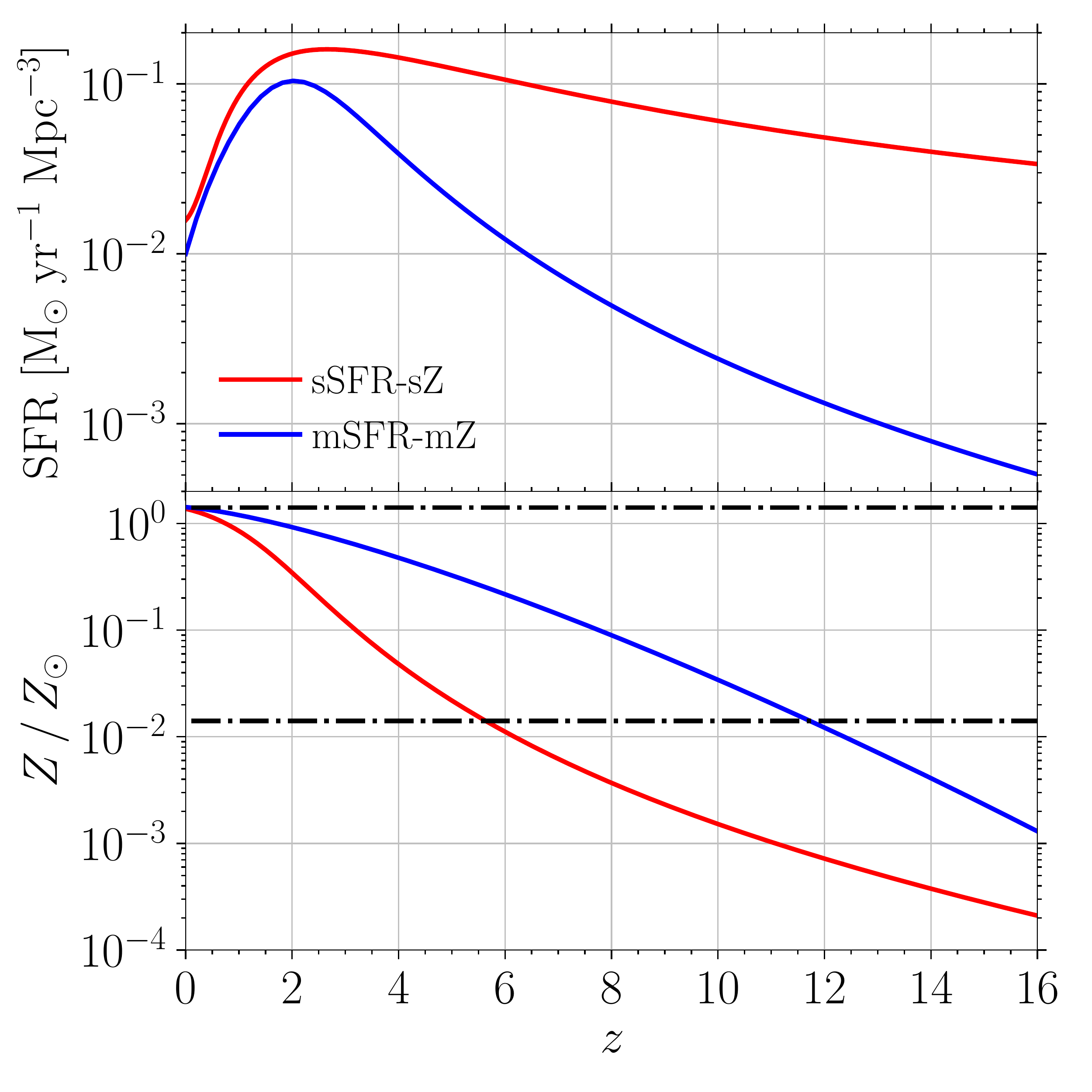}
    \caption{SFR and mean metallicity of the galaxy population $\langle Z\rangle$ (in units of solar metallicity Z$_\odot = 0.0142$) computed for the models mSFR-mZ (blue lines) and sSFR-sZ (red lines) as function of redshift. The dashed-dotted lines in the lower panel represent the range of metallicity that SEVN can evolve. Stars with metallicity exceeding our range are treated as stars in the lowest/highest metallicity bin.}
    \label{fig:SFR_and_metal}
\end{figure}

We consider two main models for the star formation rate (SFR) and the evolution of the mean metallicity $\langle Z\rangle$ of the galaxy population across cosmic history. The first model takes both from \citet{Madau_2017} and is labelled as ``mSFR-mZ''. The second model adopts the SFR as in \citet{Strolger_2004} and the metallicity for the intergalactic medium reported in \citet{Madau_2014}, shifted to match \citet{Madau_2017} metallicity at $z = 0$. This accounts for a possible rapid decline of the metallicity between the present and redshfit $z \simeq 4$. The model is labelled as ``sSFR-sZ''. The SFR and mean metallicity $\langle Z\rangle$ versus redshift are shown in Fig.~\ref{fig:SFR_and_metal} for these two models. We also consider two additional intermediate models, combining the SFRs and the metallicity prescriptions, thus labelled ``mSFR-sZ'' and ``sSFR-mZ'' (see also \citealt{10.1093/mnras/sty3087, Neijssel2019} for further discussions about uncertainties on the SFR and metallicity distribution).

We assume a stellar initial mass function (IMF) $\xi(M_\star, \alpha)\propto M_\star^{-\alpha}$ between $[8, \, 350] \msun$, with $\alpha = 2.7$ for the SFR from \citealt{Madau_2017} and $\alpha = 2.35$ for the SFR from \citealt{Strolger_2004}. The differential  comoving volume number density formation rate of progenitor stars as a function of cosmic time, mass and redshift is
\begin{equation}
     \label{eq:density_prog_mass_metal}
     \frac{\ud^3n_\star}{\ud t \ud \log M_\star \, \ud \log Z } =  \frac{f_\star (\alpha)}{\langle m_\star(\alpha)\rangle} {\rm SFR(\emph{t})} \times p( \log M_\star) \times p( \log Z).
\end{equation}
Here $f_\star $ and $m_\star$ are the fraction of simulated binaries and the average IMF mass as defined by Equation (6) and (7) of \cite{PhysRevD.84.124037}, SFR is the cosmic star formation rate density at cosmic time $t$, and $p( \log M_\star)$, $p( \log Z)$ are the probability densities of stellar mass and metallicity. The former is directly proportional to the IMF, while the latter is taken at each redshift to be a log-normal distribution centred along either the ``mZ'' or ``sZ'' relations (as shown in Figure \ref{fig:SFR_and_metal}) with dispersion $0.5$ dex. For a given interval $(\Delta \log M_\star, \, \Delta \log Z)$, we evolve a single star with SEVN to determine its BH final mass. In this way, Equation \eqref{eq:density_prog_mass_metal} is mapped into the relic BH formation rate density, $\ud^2n/(\ud t\, \ud M_1)$. The primary BH mass, $M_1$, of each BHB is drawn from this distribution. 

To convert the formation rate of BHs into the merger rate of BHBs we make two simple assumptions: i) all BHs are in binaries with secondary BH drawn according to a flat mass ratio $q=M_{2}/M_{1}<1$ distribution in the range $[0.1, \, 1]$\footnote{Binaries with $M_2$ falling in the mass gap are discarded and the remaining population is re-normalized to match the total mass density of BHB produced to be equal to $\int {\ud \log M_1}\,M_1[\ud^2n/(\ud t\, \ud \log M_1)]$.}, and ii) mergers occur at a time $t_m=t+\tau$ \footnote{Here $\tau$ includes the evolution time of the primary star.} where the delay time $\tau$ is distributed according to $p(\tau) \propto \tau^{-1}$  ) \citep{Dominik_2012, Neijssel2019}
between  $\tau_{\rm \, min} = 50 \megayr$  and $\tau_{\rm \, max} = t_{\rm Hubble}$ \citep{2016MNRAS.461.3877D}, where $t_{\rm Hubble}$ is the Hubble time. We also explore the possibility for a flat mass ratio distribution in the range [0.5,1] \citep{2019arXiv190512669F}. The rate density  per comoving volume of merging BHBs is therefore given by:
\begin{equation}
\begin{split}
     \label{eq:density_BHB}
      &\frac{\ud^3n}{\ud t_m\, \ud \log M_{1}\,\ud q} =  \\ & {\cal C}\,\int_{t<t_m}\int\frac{\ud^4n}{\ud t\,\ud \tau\,\ud \log M_{1}\,\ud q}  \delta(t_m-(t+\tau)) \ud \tau \ud t.
\end{split}
\end{equation}
The normalization constant ${\cal C}$ is set to ensure that the intrinsic BHB merger rate in the local Universe is
\begin{equation}
\int_{5\msun}^{50\msun}\ud \log M_{1}\int\ud q\frac{\ud^3n}{\ud t_m\,\ud \log M_{1}\,\ud q}\bigg{|}_{z=0}=50 \, {\rm Gpc}^{-3}\,{\rm yr}^{-1},
\end{equation}
 close to the best estimate provided by the LIGO-Virgo O2 run  \citep{2018arXiv181112940T}. This a posteriori normalization is needed because of the very simplistic assumptions made above. We checked, however, that both the resulting BHB merger rate density as a function of redshift and the mass function of merging BHBs below the pair-instability gap are in good agreement with sophisticated population synthesis models found in the literature  \citep[e.g.][]{Spera2019} . 

 We are here interested in binaries with at least one BH above the pair-instability gap. We define their merger rate as 
\begin{equation}
\mathcal{R}(z_m)=\int_{120\msun}^{\infty}\ud \log M_{1}\int\ud q\frac{\ud^3n}{\ud t_m\,\ud \log M_{1}\,\ud q}.
\end{equation}
Depending on $q$, the secondary can be either below or above the mass gap, thus defining the sub-classes of `across-gap' and `above-gap' BHBs introduced above. The number of detections per year is then computed as
\begin{equation}\label{eq:detected_rate}
    \mathcal{R}_{\rm det} = \int \mathcal{R}(z_m) \frac{1}{1+z_{m}} \frac{\ud V_{c}}{\ud z_{m}}  p_{
    \rm det} \, \ud z_{m} 
\end{equation}
where $(1+z_{m})^{-1} = \ud t_{m}/\ud t_m^{\rm obs}$ accounts for the time dilation between the source and the observer frames and $\ud V_{c}/\ud z_{m}$ is the differential comoving volume shell. Finally, $p_{\rm det}$ represents the detection probability of a random-oriented binary with a given $M_1$, $q$ and $z_m$ for a threshold signal-to-noise ratio (S/N) \citep{Abadie_2010}.

Although Equation \eqref{eq:detected_rate} is appropriate for the detection rate of ground-based interferometers, LISA will also see persistent sources, caught several years before coalescence. The distribution of observed sources in the LISA band is simply given by

\begin{equation}
  \label{eq:dndf}
  \frac{\ud N}{\ud \log M_{1} \, \ud q \, \ud z \, \ud \ln f_{\rm gw}} = \frac{\ud^3n}{\ud t_m\, \ud \log M_{1}\,\ud q} \frac{\ud V_{\rm c}}{\ud z}\frac{\ud t_m}{\ud \ln f_{\rm gw}},
\end{equation}
where $\ud t_m/\ud \ln f_{\rm gw}$ is given by the quadrupole approximation for circular orbit \citep{PhysRev.136.B1224} as
\begin{equation}
\label{eq:dfdt}
    \dfrac{\ud t_m}{\ud \ln f_{\rm gw}} = \dfrac{5}{96 \pi^{8/3}} \left (\dfrac{c^3}{G \mathcal{M}} \right)^{5/3} (f_{\rm gw} (1+z))^{-8/3}.
\end{equation}
Here $\mathcal{M}=(M_{1}M_{2})^{3/5}/(M_{1}+M_{2})^{1/5}$ is the source frame chirp mass and $f_{\rm gw}$ the observed GW frequency \footnote{We assume circular BHBs because several processes acting during stellar evolution (e.g. tidal circularisation, common envelope evolution, etc.) and long delay times are expected to produce nearly circular BHB orbits.}.
For each BHB population model  described at the beginning of this section, equation \eqref{eq:dndf} is used to draw 10 Monte Carlo  realisations of the BHB distribution across the observed frequency spectrum  in the range $[10^{-4}, \, 10^{-1}]\,  \rm Hz$. Each sample is then taken to represent the distribution of sources in the sky at the start of the LISA mission.

\begin{figure*}
	\includegraphics[width=\textwidth]{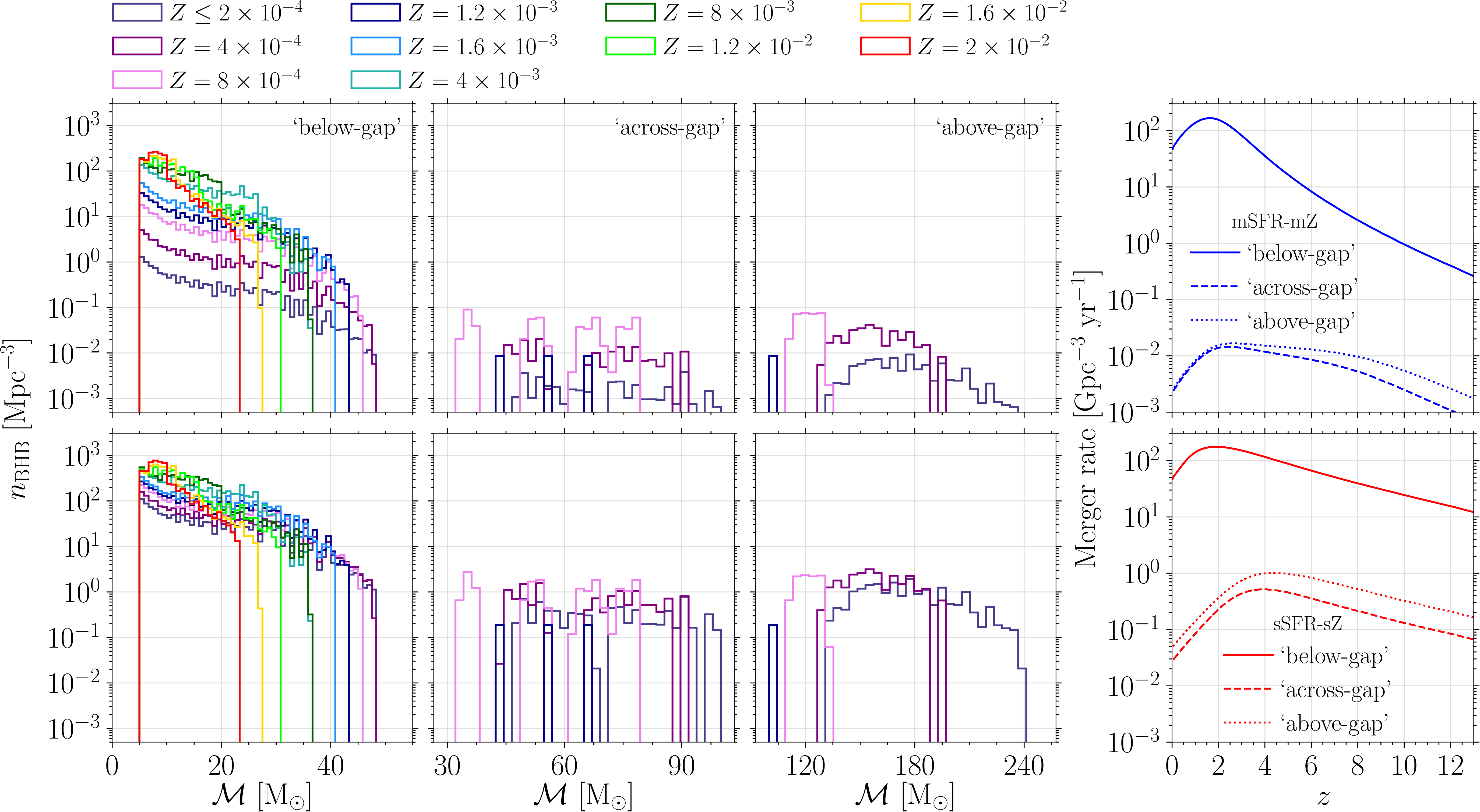}
    \caption{Number density of BHBs formed per unit comoving volume (in units of $ \Mpcmcube$) versus $\mathcal{M}$ for different values of the metallicity (six left-most panels) and merger rate per comoving $\Gpcmcube$ versus redshift (two right-most panels).
      {\it Left-most panels}: binaries are divided (from left to right) as `below-gap' binaries, `across-gap' binaries and `above-gap' binaries.
    {\it Right-most panels}: the merger rate density is broken down into the three BHB sub-populations: `below-gap' (solid lines), `across-gap' (dashed lined) and `above-gap' (dotted lined).
    {\it Upper panels}: model mSFR-mZ. {\it Lower panels}: model sSFR-sZ.}
    \label{fig:mchirp_formed}
\end{figure*}

All sources are then evolved forward in time assuming GW-driven dynamics and their S/N in the detector is evaluated as

\begin{equation}\label{eq:SNR_GWB}
    ({\rm S/N})^2 = \int_{}^{} \frac{ |\tilde{h}(f_{\rm gw}, M_{1}, q, z)|^2 }{S_{\rm n}(f_{\rm gw})} \ud f_{\rm gw},
\end{equation}
where $\tilde{h}(f_{\rm gw})$ is the Fourier transform of the GW strain and $S_{\rm n}(f_{\rm gw})$ is the power spectral density of the detector~\footnote{For O2 LIGO sensitivity, we adopt the curve labelled as 'mid' in \citealt{Abbott2016}.  
We also consider Advanced LIGO \citep[aLIGO,][]{2010CQGra..27h4006H}, Advanced Virgo \citep[AdVirgo,][]{Acernese_2014} and Einstein Telescope \citep[ET-D,][]{2011CQGra..28i4013H}. For LISA we adopt the curve described in \citealt{Robson_2019}.}. Note that the integral in Eq.~\eqref{eq:SNR_GWB} is over the frequency interval covered by the source over the observation time. For each value of $M_{1},q$ and $z_{\rm m}$, we compute the S/N randomizing over sky-position, polarization and inclination angles, and assuming non-spinning BHs. For ground-based detectors, we compute the S/N with the LALsuite \citep{lalsuite}. We model the inspiral-merger-ringdown signal with the IMRPhenomD waveform \citep{PhysRevD.93.044006, PhysRevD.93.044007}. For O1/O2, we consider an event to be detectable if $\rm S/N > 8$ \citep{LIGOScientific:2018mvr}, while for LIGO/Virgo at design sensitivity and ET we assume $\rm S/N > 12$. Similarly, for LISA we compute the S/N with the IMRPhenomC waveform \citep{PhysRevD.82.064016} with $\rm S/N > 8$. To estimate the rates for multiband events, we consider only the events detected in LISA and coalescing in   $T_{ \rm gw} < 4 \yr$, where $T_{\rm gw}$ is the merger timescale due to GW emission. We also consider a possible extended time mission of 10 years.
 
We compute the level of the stochastic GWB generated by the inspiralling BHBs, at each frequency by summing in quadrature the characteristic strains of all unresolved sources, i.e. binaries with $\rm S/N < 8$. Then, the signal power S/N$_{\rm gwb}$ is evaluated following \citet{Thrane2013} and \citet{Sesana2016}

\begin{equation}
    ({\rm S/N_{\rm gwb}})^2 = T \int \gamma(f_{\rm gw}) \dfrac{h_{c,\rm gwb}^4}{f_{\rm gw}^2 S_n(f_{\rm gw})^2} \ud f_{\rm gw}
\end{equation}
where $T = 4$ yr is the LISA mission required lifetime, $h^2_{c, \rm gwb}(f_{\rm gw})  = 2 f_{\rm gw} S_h(f_{\rm gw})$ (being $S_h(f_{\rm gw})$ the power spectral density of the signal), and $\gamma(f_{\rm gw})=1$ \citep[see Fig.~4 in][]{Thrane2013}.
We estimate the strength of the GWB through its GW energy density parameter
\begin{equation}
  \label{eq:omegagw}
   \Omega_{\rm gw}(f_{\rm gw}) = \dfrac{2}{3} \left(\dfrac{\pi f_{\rm gw} h_{c, \rm gwb}}{H_0}\right)^2,
\end{equation}
where $H_0$ the Hubble's constant. 

\section{Populations, Rates and GW Background}
\label{sec:pop_and_rate} 

In Fig.~\ref{fig:mchirp_formed} we show the number density of BHBs formed per unit comoving volume versus $\mathcal{M}$, for different metallicities and sub-populations, and the corresponding merger rate for models mSFR-mZ and sSFR-sZ.
Given the distribution of the mass ratio adopted, there is no evident gap in the source-frame chirp mass, and the `across-gap' and `above-gap' sub-populations are the result of poor-metal stars with $\langle Z \rangle < 1.2 \times 10^{-3}$. For the `below-gap' sub-populations, the outcome of this analysis is fairly consistent with that of \citet{Spera2019} (see their Fig.~D2). At $ \langle Z \rangle > 8 \times 10^{-3}$ our maximum chirp mass is close to theirs, while at $\langle Z \rangle < 8 \times 10^{-3}$, we obtain larger chirp masses, in the range $[40, \, 50] \msun$. This is expected due to the difference from single to binary evolution. However we note that $\mathcal{M} \lesssim 50 \msun$ have been recovered in alternative population synthesis codes \citep{10.1093/mnras/sty3087}.

The comparison between the two models shown in Fig.~\ref{fig:mchirp_formed} highlights the impact of metallicity on the number density of heavy BHBs. Model sSFR-sZ predicts a rapid decline in the metallicity versus redshift and, as a consequence, the `across-gap' and `above-gap' sub-population rates are one order of magnitude higher than in model  mSFR-mZ. This is also evident in the right panels showing the merger rate density of the three sub-populations for each model. Note that in both models, the `above-gap' sub-population produces slightly more mergers than the `across-gap' one, but the total merger rate is always heavily dominated by the `below-gap' systems. Mixed models ( mSFR-sZ and sSFR-mZ, not shown) give intermediate results, as expected.

\begin{figure}
    \includegraphics[width=0.5\textwidth]{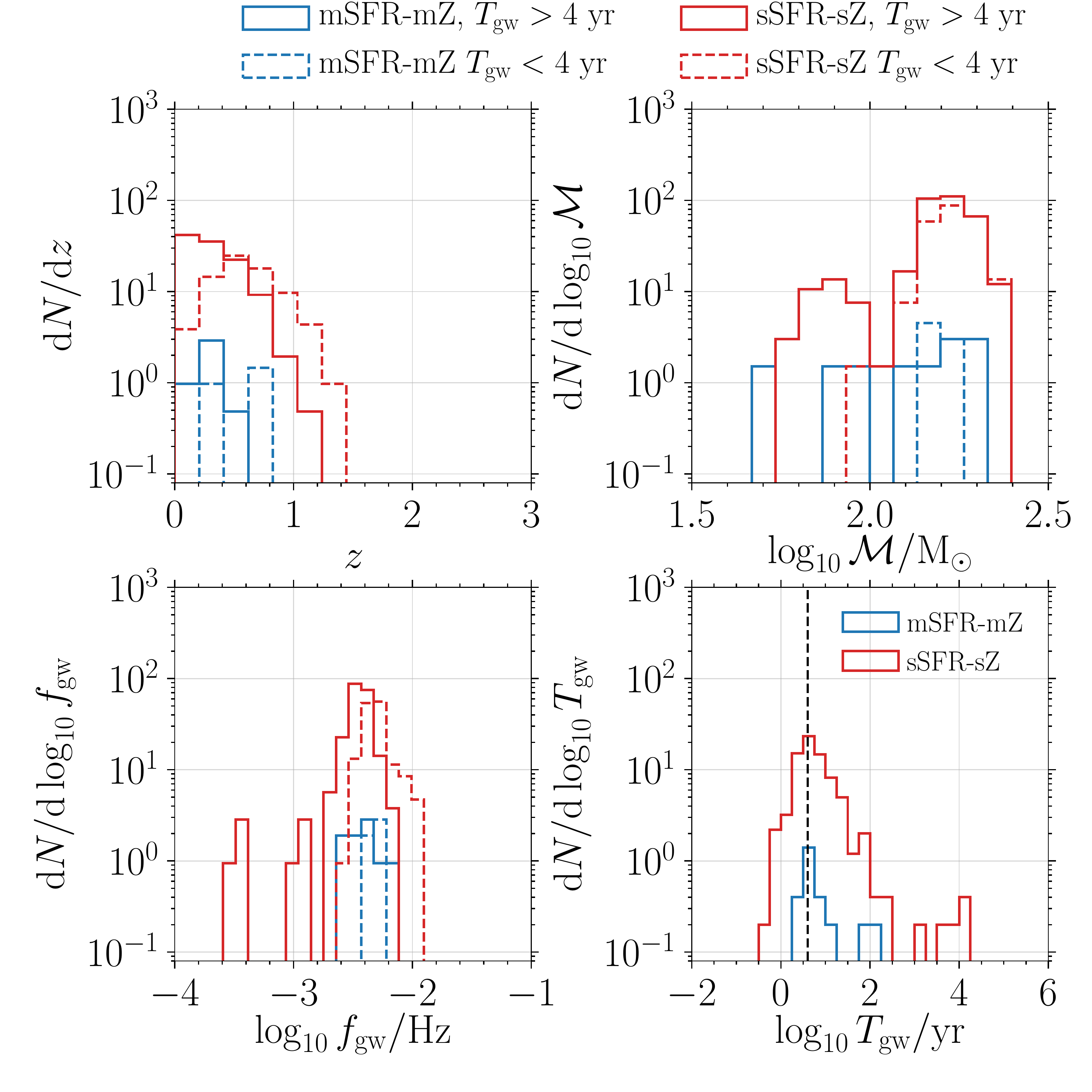}
    \caption{Differential number of events with LISA ${\rm S/N} > 8$ assuming 4 years of observations and models mSFR-mZ (blue) and sSFR-sZ (red) as a function of redshift, rest-frame chirp mass, observed GW frequency and time to coalescence. Colour code and line style are labelled in the figure.}
    \label{fig:sampling_mmdd}
\end{figure}
\begin{figure}
    \includegraphics[width=0.5\textwidth]{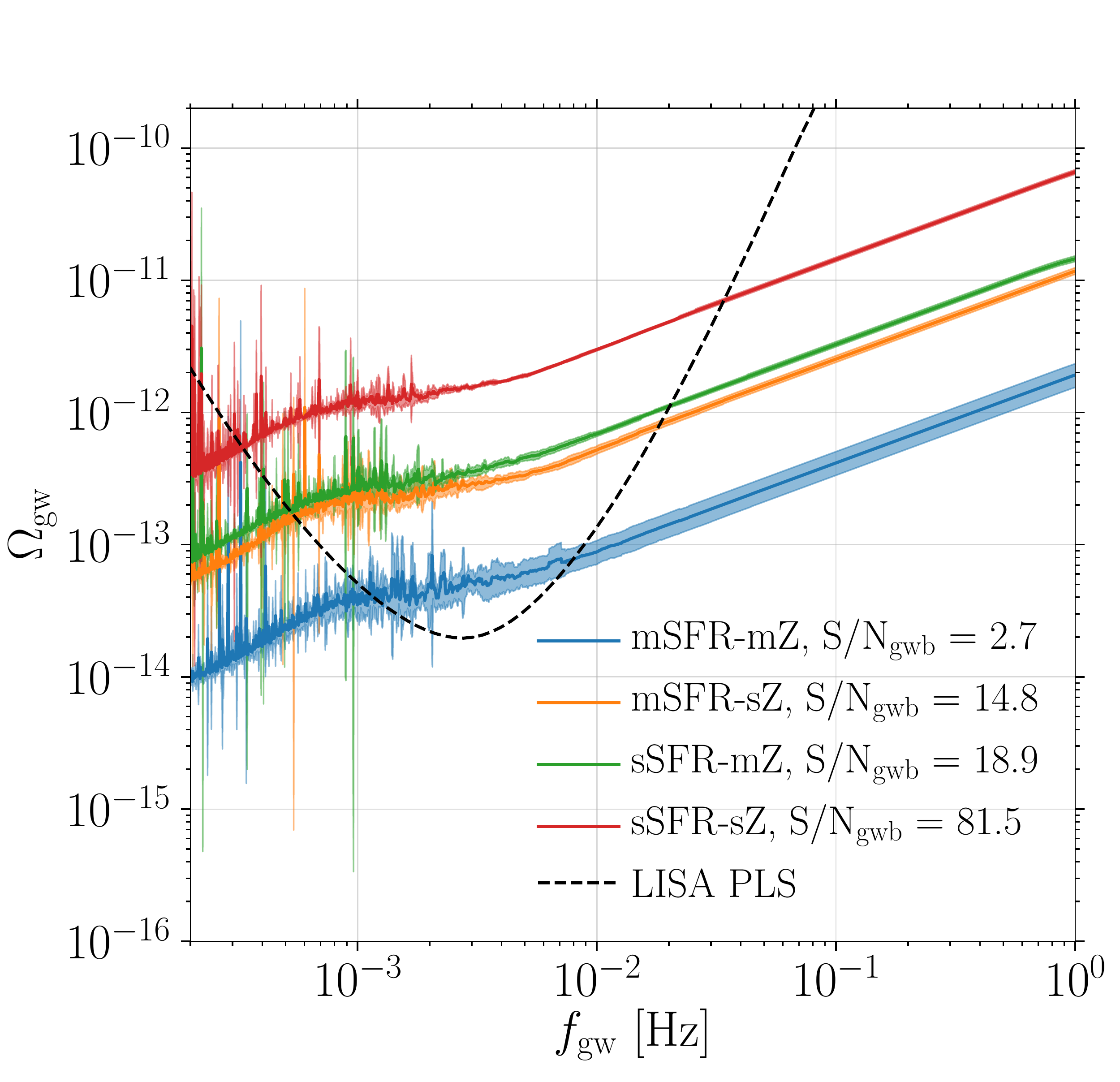}
    \caption{GW energy density parameter for the four explored models as labelled. The dashed black curve represents the LISA power-law sensitivity (PLS) curve adapted from \citep{Caprini2019} assuming a threshold of ${\rm S/N}_{\rm gwb} = 1$.}
    \label{fig:omega_gw}
\end{figure}

\begin{table*}
\begin{ruledtabular}
\begin{tabular}{l| c c c c c c c c } 
        & \multicolumn{8}{c}{Models}\\
 Detection rates  & \multicolumn{2}{c}{mSFR-mZ}  & \multicolumn{2}{c}{sSFR-sZ} & \multicolumn{2}{c}{sSFR-mZ} & \multicolumn{2}{c}{mSFR-sZ} \\ 
 
 & `across' & `above' & `across' & `above' & `across' & `above' & `across' & `above' \\

\Xhline{0.1pt} \\
Rate O1/O2 $({\rm S/N} > 8$) $[\yrmone]$ &	$0.001$	&  $0.001 $	&	$0.01$	&	$0.03$	&	$0.005$  &  $0.007$  &  $0.003$  &  $0.005$ \\ 
				
Rate LIGO/Virgo design $({\rm S/N} > 12$) $[\yrmone]$ &	$0.1$	&  $0.4 $	&	$0.9$	&	$6.9$	&	$0.4$  &  $1.9$  &  $0.3$  &  $1.6$ \\ 

Rate ET $({\rm S/N} > 12$) $[\yrmone]$ &	$8.1$	&  $10.7$	&	$212.8$	&	$458.5$	&	$61.7$  &  $116.3$  &  $39.8$  &  $68.2$ \\ 

\hline\hline\\

Detected events in 4 yr  & \multicolumn{8}{c}{}\\

\hline\\
LISA events (${\rm S/N} > 8$) &	$0.2$	&  $1.3$	&	$2.9$	&	$42.5$	&	$0.6$  &  $12.9$  &  $0.3$  &  $8.6$\\ 

LISA events (${\rm S/N} > 8$ \& $T_{\rm gw} < 4 \yr$) &	$<0.1$	&  $0.8$	&	$0.5$	&	$19.3$	& $<0.1$  & $5.8$  &  $<0.1$  &  $4.6$  \\

LIGO/Virgo multiband events (${\rm S/N} > 12$) &	$<0.1$	&  $0.5$		&	$0.4$		&	$8.8$		&	$<0.1$	  &  $2.9$	  &  $<0.1$	  &  $1.9$	 \\ 

ET multiband events (${\rm S/N} > 12$) &	$<0.1$		& $0.8$		& $0.5$		&	$19.3$		&	$<0.1$	  &  $5.8$	  &  $<0.1$	  &  $4.6$	 \\ 

S/N$_{\rm gwb}$ background in LISA & \multicolumn{2}{c}{2.7}  & \multicolumn{2}{c}{81.5} & \multicolumn{2}{c}{18.9} & \multicolumn{2}{c}{14.8}\\

\hline\hline\\
Detected events in 10 yr  & \multicolumn{8}{c}{}\\

\hline\\
LISA events (${\rm S/N} > 8$) &	$0.6$	&  $6.2$	&	$6.2$	&	$152.1$	&	$1.9$  &  $33.0$  &  $1.7$  &  $27.0$\\ 

LISA events (${\rm S/N} > 8$ \& $T_{\rm gw} < 10 \yr$) &	$0.1$	&  $3.6$	&	$1.5$	&	$102.6$	& $0.3$  & $22.1$  &  $0.6$  &  $18.2$  \\

LIGO/Virgo multiband events (${\rm S/N} > 12$) &	$1$	&  $1.6$		&	$1.2$		&	$34.7$		&	$0.2$	  &  $8.6$	  &  $0.5$	  &  $5.6$	 \\ 

ET multiband events (${\rm S/N} > 12$) &	$0.1$		& $3.6$		& $1.5$		&	$102.6$		&	$0.3$	  &  $22.1$	  &  $0.6$	  &  $18.2$	 \\ 

S/N$_{\rm gwb}$ background in LISA & \multicolumn{2}{c}{3.7}  & \multicolumn{2}{c}{117.8} & \multicolumn{2}{c}{26.9} & \multicolumn{2}{c}{21.4}\\

\end{tabular}
\end{ruledtabular}
\caption{\label{tab:summary_tab}Upper panel: Number of mergers detected per year by O1/O2,  LIGO/Virgo at design sensitivity and ET for our four different models as labelled in the text. Lower panel: Number of event over four years for LISA, number of events merging within four years to unable multiband observation with LIGO/Virgo at design sensitivity and with ET in case of joint observations. Last row gives the SNR from the stochastic GW background (summing `across-gap' and `above-gap' sub-populations) for the four models. For each model, left (right) column refers to `across-gap' (`above-gap') binaries.}
\end{table*}

In Fig.~\ref{fig:sampling_mmdd} we show the overall properties of the events observed by LISA with  $\rm S/N > 8$. The majority of the events concentrates at $z \lesssim 0.5$ with the tail extending up to $z \approx 1.5$ in model sSFR-sZ. In addition, for both models, systems merging within the LISA lifetime (dashed lines) are detected at slightly higher redshifts. The source-frame chirp mass distributions cluster around two peaks that broadly correspond to the `across-gap' (left peak) and `above-gap' (right peak) sub-populations. It is also evident that the event number in the `above-gap' group is nearly $\times10$ higher than the `across-gap' one. This is  simply because  more massive and nearly equal-mass binaries produce louder GW signal and can be seen further. Note that, in these models, we do not expect any detected `across-gap' binaries merging in 4 years. Most of the detected sources have an initial $f_{\rm gw}$ around $\sim 3\times 10^{-3}$ Hz, while only for model sSFR-sZ we can observe around $\sim \mathcal{O}(1)$ sources down to $10^{-4}$ Hz. Obviously, merging BHBs (dashed lines) peak at a slightly higher frequencies with respect to persistent ones (solid lines) for both models.

Fig.~\ref{fig:omega_gw} shows the energy density of the stochastic GWB -- $\Omega_{\rm gw}$, of Equation \eqref{eq:omegagw}-- as a function of observed frequency for all the considered models. Except for model mSFR-mZ, the signal is strong enough to be separated from the LISA detector noise. In the most optimistic model, $\Omega_{\rm gw}$ is comparable to that inferred from the `below-gap' BHB and neutron star binary populations \citep[e.g.][]{2003MNRAS.346.1197F,Sesana2016}, thus significantly contributing to the family of unresolved LISA astrophysical foregrounds. Our result is also consistent with the upper limits reported for O1/O2 from the LIGO/Virgo collaboration \citep{PhysRevLett.118.121101, 2019arXiv190302886T}.  

Besides posing another compelling scientific case for LISA, the population of `above-gap' BHBs is also important for ground-based detectors. This is quantified in Tab. \ref{tab:summary_tab}, summarizing all the relevant figures of this study.  The first three rows of the table report the merger rates, computed according to Eq.~\eqref{eq:detected_rate} for LIGO/Virgo at O2, design sensitivity and for ET. Current non detection of `across-gap' and `above-gap' binaries during 01/02 is consistent with our models. LIGO/Virgo at design sensitivity can detect between $\simeq 0.5 $ and $\simeq 7$ `above-gap' events per year for the pessimistic and optimistic models, respectively. Intermediate models predict $\approx 2$ events per year. For ET, the rate increases by more than an order of magnitude for all sub-populations, independently of the model. The number of detected events range from $ \simeq [10, \,  460] \,  \yrmone$. Due to the improved sensitivity, ET would also be able to detect several `across-gap' binaries per year in all models.

For a 4 year LISA mission, the number of detected `above-gap' events ranges between $\simeq 1.5$ and $43$ for the two limiting models. The number of events detected with $\rm S/N >8$ and merging in the mission lifetime is reduced but it's still of order unity even in the pessimistic model. In particular, the case sSFR-sZ predicts $ \simeq 20$ events, while the two intermediate models predict around $\simeq 5$ events. These sources are the best candidate for multiband detection. If LISA will join LIGO/Virgo, we expect to detect from $\simeq 0.5$ to $\simeq 10$ multiband binaries. If ET will be operative instead, all the sources detected in LISA  will also be detected at later times by ET. If LISA reaches the 10 years mission goal, the number of detected events increase by a factor of $\gtrsim 3$. This can be explained considering that the increase in SNR from a longer time mission goes as $\sim \sqrt{10/4}$, which translates in almost a factor of four in the accessible volume and, therefore, in the number of sources, assuming a constant merger rate.

We also test the case for a flat mass ratio distribution for $q \in [0.5,\, 1]$, which seems to be favoured by current LIGO-Virgo detections \citep{LIGOScientific:2018mvr}. Under this assumption, we lose the `across-gap' sub-populations, since `across-gap' systems would necessarily have $q<0.5$, but we find a significant increase of $\approx2$ in the rate for the `above-gap' sub-population. 

Finally, we run a Fisher Matrix code to estimate the uncertainties in the parameter estimation of these populations in the LISA band. The parameter estimation is performed with an 8 post-Newtonian (8PN) frequency domain waveform for circular precessing BHs \citep{PhysRevD.90.124029}. Due to the larger number of cycles in band and to LISA orbital motion, we expect to be able to localize multiband sources with a median precision of $\lesssim 10 \, \rm deg^2$ and determine single component redshifted masses to better than $1\%$ percent. The luminosity distance is determined with a median uncertainty of  $\lesssim 30\%$, while the spins would be essentially unresolved.

\section{Conclusion}
\label{sec:conclusion}

The existence of stellar BH above the pair-instability gap is an uncharted territory, as there are many uncertainties on the formation and evolution of very massive stars, on the gravitational collapse and nuclear energy production, on  binary formation in galactic fields and dynamical systems. Any detection of BHBs belonging to this population will be of capital importance in the understanding of the extreme physics governing the evolution of massive stars. In this Letter we performed a first attempt to quantify the population of `above-gap' BHBs observable with current and future ground- and space-based detectors.  Since it works under the ansatz that `above-gap' BHBs form and evolve abiding the same physics governing the evolution of `below-gap' systems, our approach can be considered agnostic. As such, although employed for the field formation scenario, it can be in principle extended to alternative formation channels (e.g. dynamical capture), where massive BHs form from multiple GW-driven mergers \citep{PhysRevD.100.043027}.

We find that prospects for discovering `above-gap' BHBs are interesting, with several systems detectable either from the ground or from space (see numbers in Tab. \ref{tab:summary_tab}). Moreover, `above-gap' binaries are primary candidates for multi-band detection, with up to ${\cal O}(100)$ observable systems for a detector network featuring a 10 yr LISA mission plus 3G interferometers on the ground \citep{Kalogera2019}.  A critical ansatz underlying those numbers is the extension of the Kroupa IMF up to $350 \msun$.  The formation of such massive stars is still puzzling, although there is some observational evidence that the stellar IMF might in fact extend to $M\gtrsim 300\msun$  \citep{10.1111/j.1365-2966.2010.17167.x}. We note that if massive stars do not form, the merger rate for `above-gap' binaries from this channel has to be zero. On the other hand, future detections of `above-gap' binaries can empirically prove that massive stars form.

Finally, we stress that the numbers presented here are subject to large uncertainties, stemming from several pieces of poorly known underlying physics, including the detailed evolution of massive star, the cosmic evolution of stellar metallicity and its dependence on the galactic environment \citep{10.1093/mnras/stw505,10.1093/mnras/stx3028}, the mass and mass-ratio distribution of binaries and so on. We considered in this Letter a minimal set of models, bracketing some of the critical uncertainties. An extended study, exploring the whole parameter space is ongoing and will be the subject of a future publication. 

\acknowledgements
The authors acknowledge P. Madau, R. Schneider, C. Berry, A. Lupi  and the anonymous referee for insightful discussions and comments.
A.~M., M.~B. and M.~C. acknowledge partial financial support from the INFN TEONGRAV specific initiative.

\bibliography{bibliography.bib}

\begin{thebibliography}{}
\expandafter\ifx\csname natexlab\endcsname\relax\def\natexlab#1{#1}\fi

\bibitem[{Aasi {et~al.}(2015)}]{2015}
Aasi, J., {et~al.} 2015,
  \href{http://dx.doi.org/10.1088/0264-9381/32/7/074001}{\color{magenta}Classical
  and Quantum Gravity}, 32, 074001

\bibitem[{Abadie {et~al.}(2010)}]{Abadie_2010}
Abadie, J., {et~al.} 2010,
  \href{http://dx.doi.org/10.1088/0264-9381/27/17/173001}{\color{magenta}Classical
  and Quantum Gravity}, 27, 173001

\bibitem[{Abbott {et~al.}(2016{\natexlab{a}})}]{PhysRevLett.116.241102}
Abbott, B.~P., {et~al.} 2016{\natexlab{a}},
  \href{http://dx.doi.org/10.1103/PhysRevLett.116.241102}{\color{magenta}Phys.
  Rev. Lett.}, 116, 241102

\bibitem[{Abbott {et~al.}(2016{\natexlab{b}})Abbott, {The LIGO Scientific
  Collaboration}, {Virgo Collaboration}, Abbott, Abbott, Abernathy, Acernese,
  Ackley, Adams, Adams, Addesso, Adhikari, Adya, Affeldt, Agathos, Agatsuma,
  Aggarwal, Aguiar, Ain, Ajith, Allen, Allocca, Altin, Amariutei, Anderson,
  Anderson, Arai, Araya, Arceneaux, Areeda, Arnaud, Arun, Ashton, Ast, Aston,
  Astone, Aufmuth, Aulbert, Babak, Baker, Baldaccini, Ballardin, Ballmer,
  Barayoga, Barclay, Barish, Barker, Barone, Barr, Barsotti, Barsuglia, Barta,
  Bartlett, Bartos, Bassiri, Basti, Batch, Baune, Bavigadda, Bazzan, Behnke,
  Bejger, Belczynski, Bell, Bell, Berger, Bergman, Bergmann, Berry, Bersanetti,
  Bertolini, Betzwieser, Bhagwat, Bhandare, Bilenko, Billingsley, Birch,
  Birney, Biscans, Bisht, Bitossi, Biwer, Bizouard, Blackburn, Blair, Blair,
  Blair, Bloemen, Bock, Bodiya, Boer, Bogaert, Bogan, Bohe, Bojtos, Bond,
  Bondu, Bonnand, Bork, Boschi, Bose, Bozzi, Bradaschia, Brady, Braginsky,
  Branchesi, Brau, Briant, Brillet, Brinkmann, Brisson, Brockill, Brooks,
  Brown, Brown, Brown, Buchanan, Buikema, Bulik, Bulten, Buonanno, Buskulic,
  Buy, Byer, Cadonati, Cagnoli, Cahillane, Calder{\'o}n~Bustillo, Callister,
  Calloni, Camp, Cannon, Cao, Capano, Capocasa, Carbognani, Caride,
  Casanueva~Diaz, Casentini, Caudill, Cavagli{\`a}, Cavalier, Cavalieri, Cella,
  Cepeda, Cerboni~Baiardi, Cerretani, Cesarini, Chakraborty, Chalermsongsak,
  Chamberlin, Chan, Chao, Charlton, Chassande-Mottin, Chen, Chen, Cheng,
  Chincarini, Chiummo, Cho, Cho, Chow, Christensen, Chu, Chua, Chung, Ciani,
  Clara, Clark, Cleva, Coccia, Cohadon, Colla, Collette, Constancio, Conte,
  Conti, Cook, Corbitt, Cornish, Corsi, Cortese, Costa, Coughlin, Coughlin,
  Coulon, Countryman, Couvares, Coward, Cowart, Coyne, Coyne, Craig, Creighton,
  Cripe, Crowder, Cumming, Cunningham, Cuoco, Dal~Canton, Danilishin,
  D'Antonio, Danzmann, Darman, Dattilo, Dave, Daveloza, Davier, Davies, Daw,
  Day, DeBra, Debreczeni, Degallaix, De~Laurentis, Del{\'e}glise, Del~Pozzo,
  Denker, Dent, Dereli, Dergachev, DeRosa, De~Rosa, DeSalvo, Dhurandhar,
  D{\'i}az, Di~Fiore, Di~Giovanni, Di~Lieto, Di~Palma, Di~Virgilio, Dojcinoski,
  Dolique, Donovan, Dooley, Doravari, Douglas, Downes, Drago, Drever, Driggers,
  Du, Ducrot, Dwyer, Edo, Edwards, Effler, Eggenstein, Ehrens, Eichholz,
  Eikenberry, Engels, Essick, Etzel, Evans, Evans, Everett, Factourovich,
  Fafone, Fair, Fairhurst, Fan, Fang, Farinon, Farr, Farr, Favata, Fays,
  Fehrmann, Fejer, Ferrante, Ferreira, Ferrini, Fidecaro, Fiori, Fisher,
  Flaminio, Fletcher, Fournier, Franco, Frasca, Frasconi, Frei, Freise, Frey,
  Fricke, Fritschel, Frolov, Fulda, Fyffe, Gabbard, Gair, Gammaitoni, Gaonkar,
  Garufi, Gatto, Gaur, Gehrels, \& Gemme}]{Abbott2016}
Abbott, B.~P., {The LIGO Scientific Collaboration}, {Virgo Collaboration},
  {et~al.} 2016{\natexlab{b}},
  \href{http://dx.doi.org/10.1007/lrr-2016-1}{\color{magenta}Living Reviews in
  Relativity}, 19, 1

\bibitem[{Abbott {et~al.}(2017{\natexlab{a}})}]{PhysRevLett.118.221101}
Abbott, B.~P., {et~al.} 2017{\natexlab{a}},
  \href{http://dx.doi.org/10.1103/PhysRevLett.118.221101}{\color{magenta}Phys.
  Rev. Lett.}, 118, 221101

\bibitem[{Abbott {et~al.}(2017{\natexlab{b}})}]{PhysRevLett.118.121101}
---. 2017{\natexlab{b}},
  \href{http://dx.doi.org/10.1103/PhysRevLett.118.121101}{\color{magenta}Phys.
  Rev. Lett.}, 118, 121101

\bibitem[{Abbott {et~al.}(2018)}]{LIGOScientific:2018mvr}
---. 2018, arXiv:1811.12907

\bibitem[{Acernese {et~al.}(2014)}]{Acernese_2014}
Acernese, F., {et~al.} 2014,
  \href{http://dx.doi.org/10.1088/0264-9381/32/2/024001}{\color{magenta}Classical
  and Quantum Gravity}, 32, 024001

\bibitem[{{Amaro-Seoane} {et~al.}(2017)}]{2017arXiv170200786A}
{Amaro-Seoane}, P., {et~al.} 2017, ArXiv e-prints, arXiv:1702.00786

\bibitem[{{Belczynski } {et~al.}(2016){Belczynski }, {Heger, A.}, {Gladysz,
  W.}, {Ruiter, A. J.}, {Woosley, S.}, {Wiktorowicz, G.}, {Chen, H.-Y.},
  {Bulik, T.}, {O\'{}Shaughnessy, R.}, {Holz, D. E.}, {Fryer, C. L.}, \&
  {Berti, E.}}]{Belczynski2016}
{Belczynski }, {Heger, A.}, {Gladysz, W.}, {et~al.} 2016,
  \href{http://dx.doi.org/10.1051/0004-6361/201628980}{\color{magenta}A\&A},
  594, A97

\bibitem[{Belczynski {et~al.}(2016)Belczynski, Holz, Bulik, \&
  O'Shaughnessy}]{Belczynski:2016obo}
Belczynski, K., Holz, D.~E., Bulik, T., \& O'Shaughnessy, R. 2016,
  \href{http://dx.doi.org/10.1038/nature18322}{\color{magenta}Nature}, 534, 512

\bibitem[{{Caprini} {et~al.}(2019){Caprini}, {Figueroa}, {Flauger}, {Nardini},
  {Peloso}, {Pieroni}, {Ricciardone}, \& {Tasinato}}]{Caprini2019}
{Caprini}, C., {Figueroa}, D.~G., {Flauger}, R., {et~al.} 2019, arXiv e-prints,
  arXiv:1906.09244

\bibitem[{Chruslinska {et~al.}(2018)Chruslinska, Nelemans, \&
  Belczynski}]{10.1093/mnras/sty3087}
Chruslinska, M., Nelemans, G., \& Belczynski, K. 2018,
  \href{http://dx.doi.org/10.1093/mnras/sty3087}{\color{magenta}Monthly Notices
  of the Royal Astronomical Society}, 482, 5012

\bibitem[{Crowther {et~al.}(2010)Crowther, Schnurr, Hirschi, Yusof, Parker,
  Goodwin, \& Kassim}]{10.1111/j.1365-2966.2010.17167.x}
Crowther, P.~A., Schnurr, O., Hirschi, R., {et~al.} 2010,
  \href{http://dx.doi.org/10.1111/j.1365-2966.2010.17167.x}{\color{magenta}Monthly
  Notices of the Royal Astronomical Society}, 408, 731

\bibitem[{Di~Carlo {et~al.}(2019)Di~Carlo, Giacobbo, Mapelli, Pasquato, Spera,
  Wang, \& Haardt}]{10.1093/mnras/stz1453}
Di~Carlo, U.~N., Giacobbo, N., Mapelli, M., {et~al.} 2019,
  \href{http://dx.doi.org/10.1093/mnras/stz1453}{\color{magenta}Monthly Notices
  of the Royal Astronomical Society}, 487, 2947

\bibitem[{Dominik {et~al.}(2012)Dominik, Belczynski, Fryer, Holz, Berti, Bulik,
  Mandel, \& O'Shaughnessy}]{Dominik_2012}
Dominik, M., Belczynski, K., Fryer, C., {et~al.} 2012,
  \href{http://dx.doi.org/10.1088/0004-637x/759/1/52}{\color{magenta}The
  Astrophysical Journal}, 759, 52

\bibitem[{Dominik {et~al.}(2015)Dominik, Berti, O'Shaughnessy, Mandel,
  Belczynski, Fryer, Holz, Bulik, \& Pannarale}]{Dominik:2014yma}
Dominik, M., Berti, E., O'Shaughnessy, R., {et~al.} 2015,
  \href{http://dx.doi.org/10.1088/0004-637X/806/2/263}{\color{magenta}Astrophys.
  J.}, 806, 263

\bibitem[{{Dvorkin} {et~al.}(2016){Dvorkin}, {Vangioni}, {Silk}, {Uzan}, \&
  {Olive}}]{2016MNRAS.461.3877D}
{Dvorkin}, I., {Vangioni}, E., {Silk}, J., {Uzan}, J.-P., \& {Olive}, K.~A.
  2016, \href{http://dx.doi.org/10.1093/mnras/stw1477}{\color{magenta}\mnras},
  \href{https://ui.adsabs.harvard.edu/\#abs/2016MNRAS.461.3877D}{\color{cyan}461},
  3877

\bibitem[{{Farmer} \& {Phinney}(2003)}]{2003MNRAS.346.1197F}
{Farmer}, A.~J., \& {Phinney}, E.~S. 2003,
  \href{http://dx.doi.org/10.1111/j.1365-2966.2003.07176.x}{\color{magenta}\mnras},
  \href{https://ui.adsabs.harvard.edu/abs/2003MNRAS.346.1197F}{\color{cyan}346},
  1197

\bibitem[{Fishbach \& Holz(2017)}]{Fishbach_2017}
Fishbach, M., \& Holz, D.~E. 2017,
  \href{http://dx.doi.org/10.3847/2041-8213/aa9bf6}{\color{magenta}The
  Astrophysical Journal}, 851, L25

\bibitem[{{Fishbach} \& {Holz}(2019)}]{2019arXiv190512669F}
{Fishbach}, M., \& {Holz}, D.~E. 2019, arXiv e-prints, arXiv:1905.12669

\bibitem[{Gerosa \& Berti(2017)}]{Gerosa17}
Gerosa, D., \& Berti, E. 2017,
  \href{http://dx.doi.org/10.1103/PhysRevD.95.124046}{\color{magenta}Phys. Rev.
  D}, 95, 124046

\bibitem[{Gerosa {et~al.}(2019)Gerosa, Ma, Wong, Berti, O'Shaughnessy, Chen, \&
  Belczynski}]{Gerosa:2019dbe}
Gerosa, D., Ma, S., Wong, K. W.~K., {et~al.} 2019,
  \href{http://dx.doi.org/10.1103/PhysRevD.99.103004}{\color{magenta}Phys. Rev.
  D}, 99, 103004

\bibitem[{{Harry} \& {LIGO Scientific
  Collaboration}(2010)}]{2010CQGra..27h4006H}
{Harry}, G.~M., \& {LIGO Scientific Collaboration}. 2010,
  \href{http://dx.doi.org/10.1088/0264-9381/27/8/084006}{\color{magenta}Classical
  and Quantum Gravity},
  \href{http://adsabs.harvard.edu/abs/2010CQGra..27h4006H}{\color{cyan}27},
  084006

\bibitem[{{Heger} \& {Woosley}(2002)}]{Heger2002}
{Heger}, A., \& {Woosley}, S.~E. 2002,
  \href{http://dx.doi.org/10.1086/338487}{\color{magenta}\apj},
  \href{https://ui.adsabs.harvard.edu/abs/2002ApJ...567..532H}{\color{cyan}567},
  532

\bibitem[{{Hild} {et~al.}(2011)}]{2011CQGra..28i4013H}
{Hild}, S., {et~al.} 2011,
  \href{http://dx.doi.org/10.1088/0264-9381/28/9/094013}{\color{magenta}Classical
  and Quantum Gravity},
  \href{https://ui.adsabs.harvard.edu/abs/2011CQGra..28i4013H}{\color{cyan}28},
  094013

\bibitem[{Husa {et~al.}(2016)Husa, Khan, Hannam, P\"urrer, Ohme, Forteza, \&
  Boh\'e}]{PhysRevD.93.044006}
Husa, S., Khan, S., Hannam, M., {et~al.} 2016,
  \href{http://dx.doi.org/10.1103/PhysRevD.93.044006}{\color{magenta}Phys. Rev.
  D}, 93, 044006

\bibitem[{{Kalogera} {et~al.}(2019){Kalogera}, {Berry}, {Colpi}, {Fairhurst},
  {Justham}, {Mandel}, {Mangiagli}, {Mapelli}, {Mills}, {Sathyaprakash},
  {Schneider}, {Tauris}, \& {Valiante}}]{Kalogera2019}
{Kalogera}, V., {Berry}, C. P.~L., {Colpi}, M., {et~al.} 2019, \baas,
  \href{https://ui.adsabs.harvard.edu/abs/2019BAAS...51c.242K}{\color{cyan}51},
  242

\bibitem[{Khan {et~al.}(2016)Khan, Husa, Hannam, Ohme, P\"urrer, Forteza, \&
  Boh\'e}]{PhysRevD.93.044007}
Khan, S., Husa, S., Hannam, M., {et~al.} 2016,
  \href{http://dx.doi.org/10.1103/PhysRevD.93.044007}{\color{magenta}Phys. Rev.
  D}, 93, 044007

\bibitem[{Klein {et~al.}(2014)Klein, Cornish, \& Yunes}]{PhysRevD.90.124029}
Klein, A., Cornish, N., \& Yunes, N. 2014,
  \href{http://dx.doi.org/10.1103/PhysRevD.90.124029}{\color{magenta}Phys. Rev.
  D}, 90, 124029

\bibitem[{{LIGO Scientific Collaboration}(2018)}]{lalsuite}
{LIGO Scientific Collaboration}. 2018, {LIGO} {A}lgorithm {L}ibrary -
  {LALS}uite, free software (GPL), doi:10.7935/GT1W-FZ16

\bibitem[{{Madau} \& {Dickinson}(2014)}]{Madau_2014}
{Madau}, P., \& {Dickinson}, M. 2014,
  \href{http://dx.doi.org/10.1146/annurev-astro-081811-125615}{\color{magenta}Annual
  Review of Astronomy and Astrophysics},
  \href{https://ui.adsabs.harvard.edu/\#abs/2014ARA&A..52..415M}{\color{cyan}52},
  415

\bibitem[{{Madau} \& {Fragos}(2017)}]{Madau_2017}
{Madau}, P., \& {Fragos}, T. 2017,
  \href{http://dx.doi.org/10.3847/1538-4357/aa6af9}{\color{magenta}\apj},
  \href{https://ui.adsabs.harvard.edu/abs/2017ApJ...840...39M}{\color{cyan}840},
  39

\bibitem[{{Mapelli} {et~al.}(2019){Mapelli}, {Giacobbo}, {Santoliquido}, \&
  {Artale}}]{Mapelli2019MNRAS}
{Mapelli}, M., {Giacobbo}, N., {Santoliquido}, F., \& {Artale}, M.~C. 2019,
  \href{http://dx.doi.org/10.1093/mnras/stz1150}{\color{magenta}\mnras},
  \href{https://ui.adsabs.harvard.edu/abs/2019MNRAS.487....2M}{\color{cyan}487},
  2

\bibitem[{Marassi {et~al.}(2019)Marassi, Graziani, Ginolfi, Schneider, Mapelli,
  Spera, \& Alparone}]{10.1093/mnras/stz170}
Marassi, S., Graziani, L., Ginolfi, M., {et~al.} 2019,
  \href{http://dx.doi.org/10.1093/mnras/stz170}{\color{magenta}Monthly Notices
  of the Royal Astronomical Society}, 484, 3219

\bibitem[{Marassi {et~al.}(2011)Marassi, Schneider, Corvino, Ferrari, \&
  Zwart}]{PhysRevD.84.124037}
Marassi, S., Schneider, R., Corvino, G., Ferrari, V., \& Zwart, S.~P. 2011,
  \href{http://dx.doi.org/10.1103/PhysRevD.84.124037}{\color{magenta}Phys. Rev.
  D}, 84, 124037

\bibitem[{{Marchant} {et~al.}(2018){Marchant}, {Renzo}, {Farmer}, {Pappas},
  {Taam}, {de Mink}, \& {Kalogera}}]{Marchant2018}
{Marchant}, P., {Renzo}, M., {Farmer}, R., {et~al.} 2018, arXiv e-prints,
  arXiv:1810.13412

\bibitem[{{McKernan} {et~al.}(2019){McKernan}, {Ford}, {O'Shaughnessy}, \&
  {Wysocki}}]{2019arXiv190704356M}
{McKernan}, B., {Ford}, K.~E.~S., {O'Shaughnessy}, R., \& {Wysocki}, D. 2019,
  arXiv e-prints, arXiv:1907.04356

\bibitem[{{Neijssel} {et~al.}(2019){Neijssel}, {Vigna-G{\'o}mez}, {Stevenson},
  {Barrett}, {Gaebel}, {Broekgaarden}, {de Mink}, {Sz{\'e}csi}, {Vinciguerra},
  \& {Mandel}}]{Neijssel2019}
{Neijssel}, C.~J., {Vigna-G{\'o}mez}, A., {Stevenson}, S., {et~al.} 2019, arXiv
  e-prints, arXiv:1906.08136

\bibitem[{Peters(1964)}]{PhysRev.136.B1224}
Peters, P.~C. 1964,
  \href{http://dx.doi.org/10.1103/PhysRev.136.B1224}{\color{magenta}Phys.
  Rev.}, 136, B1224

\bibitem[{Pezzulli {et~al.}(2016)Pezzulli, Valiante, \&
  Schneider}]{10.1093/mnras/stw505}
Pezzulli, E., Valiante, R., \& Schneider, R. 2016,
  \href{http://dx.doi.org/10.1093/mnras/stw505}{\color{magenta}Monthly Notices
  of the Royal Astronomical Society}, 458, 3047

\bibitem[{Podsiadlowski {et~al.}(2003)Podsiadlowski, Rappaport, \&
  Han}]{Podsiadlowski2003}
Podsiadlowski, P., Rappaport, S., \& Han, Z. 2003,
  \href{http://dx.doi.org/10.1046/j.1365-8711.2003.06464.x}{\color{magenta}Monthly
  Notices of the Royal Astronomical Society}, 341, 385

\bibitem[{Postnov \& Yungelson(2014)}]{Postnov2014}
Postnov, K.~A., \& Yungelson, L.~R. 2014,
  \href{http://dx.doi.org/10.12942/lrr-2014-3}{\color{magenta}Living Reviews in
  Relativity}, 17, 3

\bibitem[{{Punturo} {et~al.}(2010){Punturo}, {Abernathy}, {Acernese}, {Allen},
  {Andersson}, {Arun}, {Barone}, {Barr}, {Barsuglia}, {Beker}, {Beveridge},
  {Birindelli}, {Bose}, {Bosi}, {Braccini}, {Bradaschia}, {Bulik}, {Calloni},
  {Cella}, {Chassande Mottin}, {Chelkowski}, {Chincarini}, {Clark}, {Coccia},
  {Colacino}, {Colas}, {Cumming}, {Cunningham}, {Cuoco}, {Danilishin},
  {Danzmann}, {De Luca}, {De Salvo}, {Dent}, {De Rosa}, {Di Fiore}, {Di
  Virgilio}, {Doets}, {Fafone}, {Falferi}, {Flaminio}, {Franc}, {Frasconi},
  {Freise}, {Fulda}, {Gair}, {Gemme}, {Gennai}, {Giazotto}, {Glampedakis},
  {Granata}, {Grote}, {Guidi}, {Hammond}, {Hannam}, {Harms}, {Heinert},
  {Hendry}, {Heng}, {Hennes}, {Hild}, {Hough}, {Husa}, {Huttner}, {Jones},
  {Khalili}, {Kokeyama}, {Kokkotas}, {Krishnan}, {Lorenzini}, {L{\"u}ck},
  {Majorana}, {Mandel}, {Mandic}, {Martin}, {Michel}, {Minenkov}, {Morgado},
  {Mosca}, {Mours}, {M{\"u}ller─Ebhardt}, {Murray}, {Nawrodt}, {Nelson},
  {Oshaughnessy}, {Ott}, {Palomba}, {Paoli}, {Parguez}, {Pasqualetti},
  {Passaquieti}, {Passuello}, {Pinard}, {Poggiani}, {Popolizio}, {Prato},
  {Puppo}, {Rabeling}, {Rapagnani}, {Read}, {Regimbau}, {Rehbein}, {Reid},
  {Rezzolla}, {Ricci}, {Richard}, {Rocchi}, {Rowan}, {R{\"u}diger}, {Sassolas},
  {Sathyaprakash}, {Schnabel}, {Schwarz}, {Seidel}, {Sintes}, {Somiya},
  {Speirits}, {Strain}, {Strigin}, {Sutton}, {Tarabrin}, {Th{\"u}ring}, {van
  den Brand}, {van Leewen}, {van Veggel}, {van den Broeck}, {Vecchio},
  {Veitch}, {Vetrano}, {Vicere}, {Vyatchanin}, {Willke}, {Woan}, {Wolfango}, \&
  {Yamamoto}}]{Punturo2010}
{Punturo}, M., {Abernathy}, M., {Acernese}, F., {et~al.} 2010,
  \href{http://dx.doi.org/10.1088/0264-9381/27/19/194002}{\color{magenta}Classical
  and Quantum Gravity},
  \href{https://ui.adsabs.harvard.edu/abs/2010CQGra..27s4002P}{\color{cyan}27},
  194002

\bibitem[{Robson {et~al.}(2019)Robson, Cornish, \& Liu}]{Robson_2019}
Robson, T., Cornish, N.~J., \& Liu, C. 2019,
  \href{http://dx.doi.org/10.1088/1361-6382/ab1101}{\color{magenta}Classical
  and Quantum Gravity}, 36, 105011

\bibitem[{Rodriguez {et~al.}(2018)Rodriguez, Amaro-Seoane, Chatterjee, \&
  Rasio}]{Rodriguez_2018}
Rodriguez, C.~L., Amaro-Seoane, P., Chatterjee, S., \& Rasio, F.~A. 2018,
  \href{http://dx.doi.org/10.1103/PhysRevLett.120.151101}{\color{magenta}Phys.
  Rev. Lett.}, 120, 151101

\bibitem[{Rodriguez {et~al.}(2019)Rodriguez, Zevin, Amaro-Seoane, Chatterjee,
  Kremer, Rasio, \& Ye}]{PhysRevD.100.043027}
Rodriguez, C.~L., Zevin, M., Amaro-Seoane, P., {et~al.} 2019,
  \href{http://dx.doi.org/10.1103/PhysRevD.100.043027}{\color{magenta}Phys.
  Rev. D}, 100, 043027

\bibitem[{Santamar\'{\i}a {et~al.}(2010)Santamar\'{\i}a, Ohme, Ajith,
  Br\"ugmann, Dorband, Hannam, Husa, M\"osta, Pollney, Reisswig, Robinson,
  Seiler, \& Krishnan}]{PhysRevD.82.064016}
Santamar\'{\i}a, L., Ohme, F., Ajith, P., {et~al.} 2010,
  \href{http://dx.doi.org/10.1103/PhysRevD.82.064016}{\color{magenta}Phys. Rev.
  D}, 82, 064016

\bibitem[{Schneider {et~al.}(2017)Schneider, Graziani, Marassi, Spera, Mapelli,
  Alparone, \& Bennassuti}]{10.1093/mnrasl/slx118}
Schneider, R., Graziani, L., Marassi, S., {et~al.} 2017,
  \href{http://dx.doi.org/10.1093/mnrasl/slx118}{\color{magenta}Monthly Notices
  of the Royal Astronomical Society: Letters}, 471, L105

\bibitem[{{Sesana}(2016)}]{Sesana2016}
{Sesana}, A. 2016,
  \href{http://dx.doi.org/10.1103/PhysRevLett.116.231102}{\color{magenta}\prl},
  \href{https://ui.adsabs.harvard.edu/abs/2016PhRvL.116w1102S}{\color{cyan}116},
  231102

\bibitem[{Spera \& Mapelli(2017)}]{Spera_2017}
Spera, M., \& Mapelli, M. 2017,
  \href{http://dx.doi.org/10.1093/mnras/stx1576}{\color{magenta}Mon. Not. Roy.
  Astron. Soc.}, 470, 4739

\bibitem[{Spera {et~al.}(2015)Spera, Mapelli, \& Bressan}]{Spera_2015}
Spera, M., Mapelli, M., \& Bressan, A. 2015,
  \href{http://dx.doi.org/10.1093/mnras/stv1161}{\color{magenta}Monthly Notices
  of the Royal Astronomical Society}, 451, 4086

\bibitem[{{Spera} {et~al.}(2019){Spera}, {Mapelli}, {Giacobbo}, {Trani},
  {Bressan}, \& {Costa}}]{Spera2019}
{Spera}, M., {Mapelli}, M., {Giacobbo}, N., {et~al.} 2019,
  \href{http://dx.doi.org/10.1093/mnras/stz359}{\color{magenta}\mnras},
  \href{https://ui.adsabs.harvard.edu/\#abs/2019MNRAS.485..889S}{\color{cyan}485},
  889

\bibitem[{{Stevenson} {et~al.}(2019){Stevenson}, {Sampson}, {Powell},
  {Vigna-G{\'o}mez}, {Neijssel}, {Sz{\'e}csi}, \& {Mandel}}]{Stevenson_2019}
{Stevenson}, S., {Sampson}, M., {Powell}, J., {et~al.} 2019, arXiv e-prints,
  arXiv:1904.02821

\bibitem[{Strolger {et~al.}(2004)Strolger, Riess, Dahlen, Livio, Panagia,
  Challis, Tonry, Filippenko, Chornock, Ferguson, Koekemoer, Mobasher,
  Dickinson, Giavalisco, Casertano, Hook, Bondin, Leibundgut, Nonino, Rosati,
  Spinrad, Steidel, Stern, Garnavich, Matheson, Grogin, Hornschemeier,
  Kretchmer, Laidler, Lee, Lucas, de~Mello, Moustakas, Ravindranath,
  Richardson, \& Taylor}]{Strolger_2004}
Strolger, L.-G., Riess, A.~G., Dahlen, T., {et~al.} 2004,
  \href{http://dx.doi.org/10.1086/422901}{\color{magenta}The Astrophysical
  Journal}, 613, 200

\bibitem[{Talbot \& Thrane(2018)}]{Talbot:2018cva}
Talbot, C., \& Thrane, E. 2018,
  \href{http://dx.doi.org/10.3847/1538-4357/aab34c}{\color{magenta}Astrophys.
  J.}, 856, 173

\bibitem[{{The LIGO Scientific Collaboration} {et~al.}(2018){The LIGO
  Scientific Collaboration}, {the Virgo Collaboration}, {Abbott},
  {et~al.}}]{2018arXiv181112940T}
{The LIGO Scientific Collaboration}, {the Virgo Collaboration}, {Abbott},
  {et~al.} 2018, arXiv e-prints, arXiv:1811.12940

\bibitem[{{The LIGO Scientific Collaboration} {et~al.}(2019){The LIGO
  Scientific Collaboration}, {the Virgo Collaboration},
  {et~al.}}]{2019arXiv190302886T}
{The LIGO Scientific Collaboration}, {the Virgo Collaboration}, {et~al.} 2019,
  arXiv e-prints, arXiv:1903.02886

\bibitem[{{Thrane} \& {Romano}(2013)}]{Thrane2013}
{Thrane}, E., \& {Romano}, J.~D. 2013,
  \href{http://dx.doi.org/10.1103/PhysRevD.88.124032}{\color{magenta}\prd},
  \href{https://ui.adsabs.harvard.edu/abs/2013PhRvD..88l4032T}{\color{cyan}88},
  124032

\bibitem[{{Uchida} {et~al.}(2019){Uchida}, {Shibata}, {Takahashi}, \&
  {Yoshida}}]{Uchida2019}
{Uchida}, H., {Shibata}, M., {Takahashi}, K., \& {Yoshida}, T. 2019, arXiv
  e-prints, arXiv:1901.08260

\bibitem[{Valiante {et~al.}(2017)Valiante, Schneider, Graziani, \&
  Zappacosta}]{10.1093/mnras/stx3028}
Valiante, R., Schneider, R., Graziani, L., \& Zappacosta, L. 2017,
  \href{http://dx.doi.org/10.1093/mnras/stx3028}{\color{magenta}Monthly Notices
  of the Royal Astronomical Society}, 474, 3825

\bibitem[{Valiante {et~al.}(2016)Valiante, Schneider, Volonteri, \&
  Omukai}]{10.1093/mnras/stw225}
Valiante, R., Schneider, R., Volonteri, M., \& Omukai, K. 2016,
  \href{http://dx.doi.org/10.1093/mnras/stw225}{\color{magenta}Monthly Notices
  of the Royal Astronomical Society}, 457, 3356

\bibitem[{{Volonteri}(2010)}]{Volonteri_2010}
{Volonteri}, M. 2010,
  \href{http://dx.doi.org/10.1007/s00159-010-0029-x}{\color{magenta}\aapr},
  \href{https://ui.adsabs.harvard.edu/abs/2010A&ARv..18..279V}{\color{cyan}18},
  279

\bibitem[{Woosley(2017)}]{Woosley_2017}
Woosley, S.~E. 2017,
  \href{http://dx.doi.org/10.3847/1538-4357/836/2/244}{\color{magenta}The
  Astrophysical Journal}, 836, 244

\bibitem[{{Woosley} {et~al.}(2002){Woosley}, {Heger}, \&
  {Weaver}}]{Woosley_2002}
{Woosley}, S.~E., {Heger}, A., \& {Weaver}, T.~A. 2002,
  \href{http://dx.doi.org/10.1103/RevModPhys.74.1015}{\color{magenta}Reviews of
  Modern Physics},
  \href{https://ui.adsabs.harvard.edu/abs/2002RvMP...74.1015W}{\color{cyan}74},
  1015

\bibitem[{{Yang} {et~al.}(2019){Yang}, {Bartos}, {Gayathri}, {Ford}, {Haiman},
  {Klimenko}, {Kocsis}, {M{\'a}rka}, {M{\'a}rka}, {McKernan}, \&
  {O'Shaugnessy}}]{2019arXiv190609281Y}
{Yang}, Y., {Bartos}, I., {Gayathri}, V., {et~al.} 2019, arXiv e-prints,
  arXiv:1906.09281

\bibitem[{{Yoshida} {et~al.}(2016){Yoshida}, {Umeda}, {Maeda}, \&
  {Ishii}}]{Yoshida_2016}
{Yoshida}, T., {Umeda}, H., {Maeda}, K., \& {Ishii}, T. 2016,
  \href{http://dx.doi.org/10.1093/mnras/stv3002}{\color{magenta}\mnras},
  \href{https://ui.adsabs.harvard.edu/abs/2016MNRAS.457..351Y}{\color{cyan}457},
  351

\end{thebibliography}

\end{document}